\documentstyle[11pt,newpasp,twoside,epsf]{article}
\def\edcomment#1{\iffalse\marginpar{\raggedright\sl#1\/}\else\relax\fi}
\reversemarginpar

\def\stacksymbols #1#2#3#4{\def\theguybelow{#2}
\def\verticalposition{\lower#3pt}
\def\spacingwithinsymbol{\baselineskip0pt\lineskip#4pt}
\mathrel{\mathpalette\intermediary#1}}
\def\intermediary#1#2{\verticalposition\vbox{\spacingwithinsymbol
\everycr={}\tabskip0pt
\halign{$\mathsurround0pt#1\hfil##\hfil$\crcr#2\crcr 
\theguybelow\crcr}}}

\def\lapproxeq{\stacksymbols{<}{\sim}{4}{1}}

\begin{document}
\title{The Impact  of Galaxy Formation on the Diffuse Background Radiation}
 \author{Joseph Silk}
\affil{Oxford University, NAPL, Keble Road, Oxford OX1 3RH, United Kingdom}
\author{Julien Devriendt}
\affil{Oxford University, NAPL, Keble Road, Oxford OX1 3RH, United Kingdom}

\begin{abstract}
  
The far infrared background is a sink for the hidden aspects of galaxy formation.  
At optical wavelengths, ellipticals and spheroids are old, even at $z \sim 1.$ 
Neither the luminous formation phase nor  their early evolution is seen in
the visible. We infer that ellipticals and, more generally, most spheroids must have
formed in dust-shrouded starbursts.  In this article, we show how separate tracking of
disk and spheroid star formation  enables us to infer that disks dominate
near the peak in the cosmic star formation rate at  $z \lapproxeq 2$  and in the 
diffuse ultraviolet/optical/infrared background, whereas spheroid formation dominates the 
submillimetre background. 

\end{abstract}

\section{Introduction: A Pathway to Galaxy Formation}

The theory of galaxy formation became firmly established once temperature 
anisotropies in the cosmic microwave background were discovered.  Only then 
did we have some indication of the amplitude and power spectrum of the 
primordial density fluctuations that seeded structure formation via
gravitational instability once the universe  became matter-dominated.  Adiabatic  
fluctuations in cold dark matter, with the nucleosynthesis baryon component 
of $\Omega_bh^2 \approx 0.02$ and the cold dark matter component of  
$\Omega_0 \approx 0.3,$ form the basic paradigm for structure formation, 
supplemented by the recent realisation that  $\Omega_0 + \Omega_{\Lambda} \approx 1$ 
and $\Omega_{\Lambda} - \Omega_0 \approx 0.4$ from the CMB $\Delta$T/T degree-scale 
and SNIa cosmological probes, respectively.

Inflation suggests that the fluctuations are nearly gaussian distributed, and the
problem of galaxy formation, as defined by the assembly of dark halos, is
sufficiently well-posed that it has been systematically probed via computer simulations, 
with varying degrees of success.
Highlights of the N-body simulation approach to structure formation include 
studies of the evolution of the galaxy correlation function, of the mass
function and structure of galaxy halos and galaxy clusters, and of the structure of
the intergalactic medium.  Much of this work has successfully accounted for the
observed characteristics of the observed universe.
However, not all is in good shape.  Galaxy formation to observers is the
assembly of the structural components of a galaxy.  Predictions of the properties
of the luminous components of galaxies have not met with much success.  
Disk sizes are far too small.  The ratio of luminosity to mass is too low.  
Even the detailed structure of the dark galaxy halos, as probed by high resolution 
N-body simulations, is producing results that are discrepant with observations: a density 
profile with a central cusp $\rho (r) \propto r^{-1.5}$ is predicted that is seen neither
in low surface brightness dwarf spirals, barred galaxies, the Milky Way, nor in giant 
ellipticals.  Moreover, excessive substructure is found, leading to apparent conflict 
with the observed galaxy luminosity function. 

Cosmologists have rushed in to resurrect cold dark matter with various alternatives. 
These range from particle astrophysics solutions that involve for example, dabbling 
with the weakly interacting nature of standard CDM, or astrophysical solutions, that invoke 
recourse to, for example, dynamical feedback of the dissipating baryon matter. 
While any of these solutions might eventually turn out to have some relevance,
 a more useful exercise may be  to assess the protogalaxy phase of galaxy formation.  
Improving our understanding of early galaxy evolution is not only closely coupled to
observation, but may also help clarify the nature of the protogalactic environment, the 
role of such astrophysical processes as feedback, and the possible need for more exotic 
physics.

Ideally, one would like to solve the galaxy formation problem numerically. While
 with a large N-body simulation, one can identify the sites of galaxy formation, the 
resolution required to adequately model dynamics alone is at the limit of current computing 
capabilities.   Incorporation of the gas hydrodynamics, required to study star formation, 
at adequate resolution to cover an entire galaxy, is not feasible.  The current limit of 
simulations only allows study of the formation of the first stellar mass clouds in a protogalaxy,
 with no possibility of studying either  efficiency of star formation or
 continuing fragmentation.
Yet the first
generation of stars must have involved millions of massive stars in order to produce the chemical
elements seen in the most metal-poor halo stars.

Cosmologists have therefore developed a semi-analytical approach to galaxy formation.  
This consists of constructing Monte-Carlo realisations of the merging histories 
of dark matter halos which have collapsed under their own self-gravity.
Galaxy formation is then followed by imposing certain simple rules, for example for the rate of 
star formation.  The size of a galaxy is determined by ensuring that angular momentum is 
conserved in the collapsing cloud.  The cloud forms with the small value of initial 
dimensionless angular momentum $\lambda \approx 0.05$ that is generated by tidal torques 
between neighbouring fluctuations, and attains virial equilibrium at a fraction $\lambda$ 
of the radius at maximum extent. A simple Schmidt law is used for the star formation rate 
in the resulting disk, and cold gas infall continues to feed the disk.  Spiral galaxies are 
relatively isolated galaxies that have not undergone a merger (that is, of nearly equal 
mass) within the past 10 Gyr or so.  Elliptical galaxies are assumed to form in major 
mergers on a dynamical time scale: stars are assumed to form in a starburst.

Semi-analytical galaxy formation is able to make realistic maps of the galaxy 
redshift distribution, and once the feedback parameter is carefully tweaked, the 
galaxy luminosity function and cosmic star formation history can be derived.
However, the model is not very robust, since exploration of the large parameter space for 
the gas and star formation physics is not really feasible.

It is therefore useful to consider fully analytic models of galaxy formation.  
We briefly describe such a model in section 2, detail the key procedure which allows us
to distinguish between spheroid/disk formation in section 3, and
present our results in section 4. 

\section{Basics of the Model}

\begin{figure}
\plotone{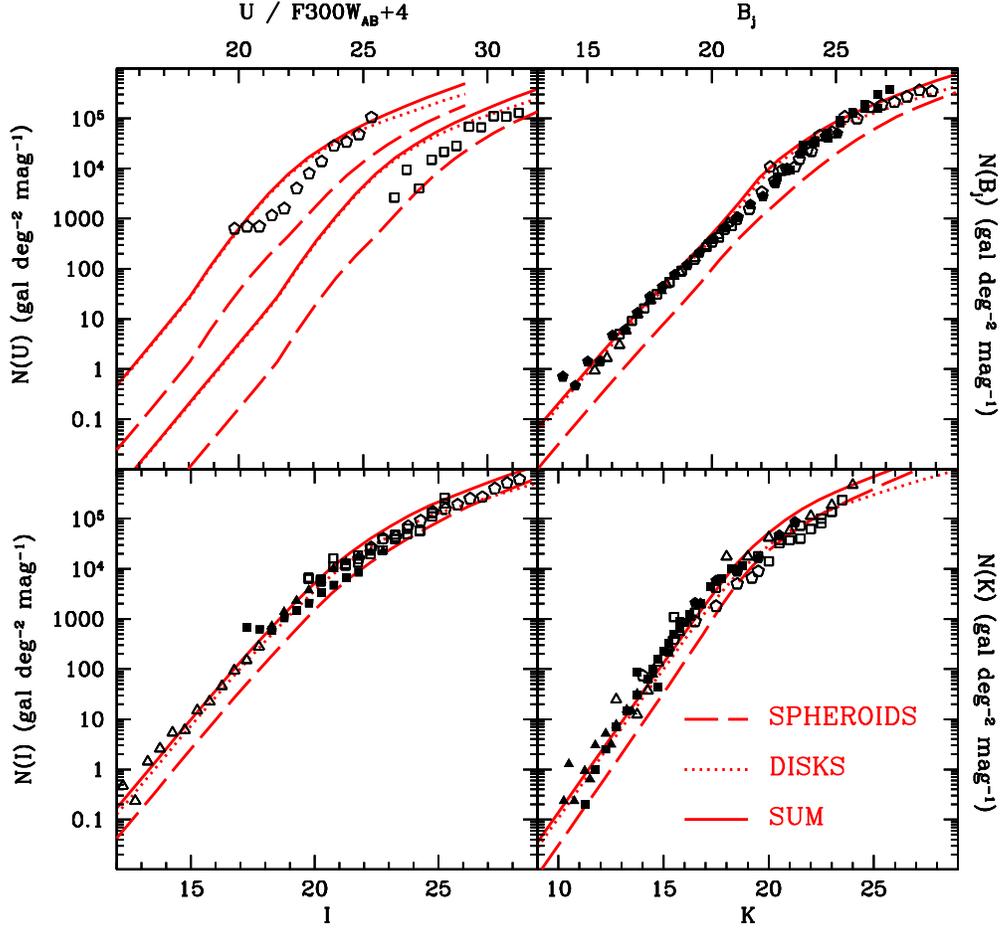}
\caption{UV/Optical/Near IR counts from our analytic model.
Coding for the lines is dotted: late type galaxies (spirals, irregulars); dashed: 
early type (ellipticals, S0s); and solid: sum of both contributions.
Data are from  Hogg et al. (1997) (U band), Williams et al. (1996) (F300W$_{\rm AB}$, 
B \& I bands), Arnouts et al. (1997) (B band), Bertin \& Dennefeld (1997) 
(B band), Gardner et al. (1996) (B, I \& K bands), Metcalfe et al. (1995)
(B band), Weir et al. (1995) (B band), Smail et al. (1995) (I band), 
Le F\`evre et al. (1995) (I band), Moustakas et al. (1997) (K band), 
and Djogorvski et al. (1995) (K band).}
\end{figure}

The cosmological framework we pick here is the canonical SCDM model 
($\Omega_0 = 1.0, \Omega_{\Lambda} =0., h = 0.5, n = 1, \sigma_8 = 0.58, \Omega_b h^2 = 0.02)$.
As discussed by Devriendt \& Guiderdoni (2000), 
only an open universe {\bf without} a cosmological constant would significantly  alter 
our conclusions, and only in the submm window. The results for a flat universe 
with a positive cosmological constant would not be quantitatively too different from 
those presented in the following. 
We then use the Press-Schechter formalism (Press \& Schechter, 1974), which allows
one to compute the comoving number of halos hosting galaxies as a function of 
redshift, provided the power spectrum of primordial fluctuations in the dark matter 
density field is known. 

To make contact with observations, star formation and chemical 
evolution must be incorporated. We follow the general framework described by 
White \& Frenk (1991), that is to say, within these halos, we let gas cool radiatively, 
settle into a disk and form stars. Technical details about the semi-empirical recipes 
used to account for the different astrophysical processes are given in 
Devriendt \& Guiderdoni (2000). Chemical and spectral evolution model are then 
computed with the {\sc stardust} model from Devriendt, Guiderdoni \& Sadat (1999). 
For sake of completeness, we mention that the dust absorption and emission model implemented in
{\sc stardust} is the 3-component model of D\'esert, Boulanger \& Puget (1990),
 which includes PAHs, very small grains and large  grains. 
This allows us to self-consistently link the optical and the far-IR/submm windows.

We emphasize that, as pointed out in Devriendt \& Guiderdoni (2000), there are 3 key 
parameters in such models, the star formation 
efficiency $\beta^{-1}$, the feedback efficiency $\epsilon_{SN}$ and the extent of 
the gaseous disks $f_c$. We take parameter  values that are  fairly similar to the ones used
by these authors, with $\beta = 60$, $\epsilon_{SN} = 0.2$ and $f_c = 4$.
These values are well within the uncertainties of observations by Kennicutt (1998) 
and Bosma (1981) for the star formation efficiency and the gaseous extent of cold disks, 
respectively. Numerical simulations by e.g. Thornton et al. (1998) tend to give values
closer to 0.1 for feedback efficiency; however our higher value is a consequence
of an attempt to reduce the number of small objects overestimated by the Press-Schechter
prescription, as well as a means to cure the ``cooling catastrophe'', where too much gas 
cools in low mass halos at high redshift.  
The qualitative effect of each these parameters can be summed up in 
the following way:
\begin{itemize}
\item  Increasing $\beta$ decreases the normalisation and faint-end slope of 
the optical and IR counts, because star formation is reduced and takes place at lower redshift.
\item Increasing $\epsilon_{SN}$ decreases the normalisation and faint-end slope of the 
optical and IR counts by quenching star formation in higher and higher mass galaxies that form at
lower and lower redshifts on average.  
\item Increasing $f_c$ means increasing the 
normalisation of the optical counts, and decreasing that of the IR counts, since it is 
is equivalent to reducing extinction.    
\end{itemize}

\begin{figure}
\plotone{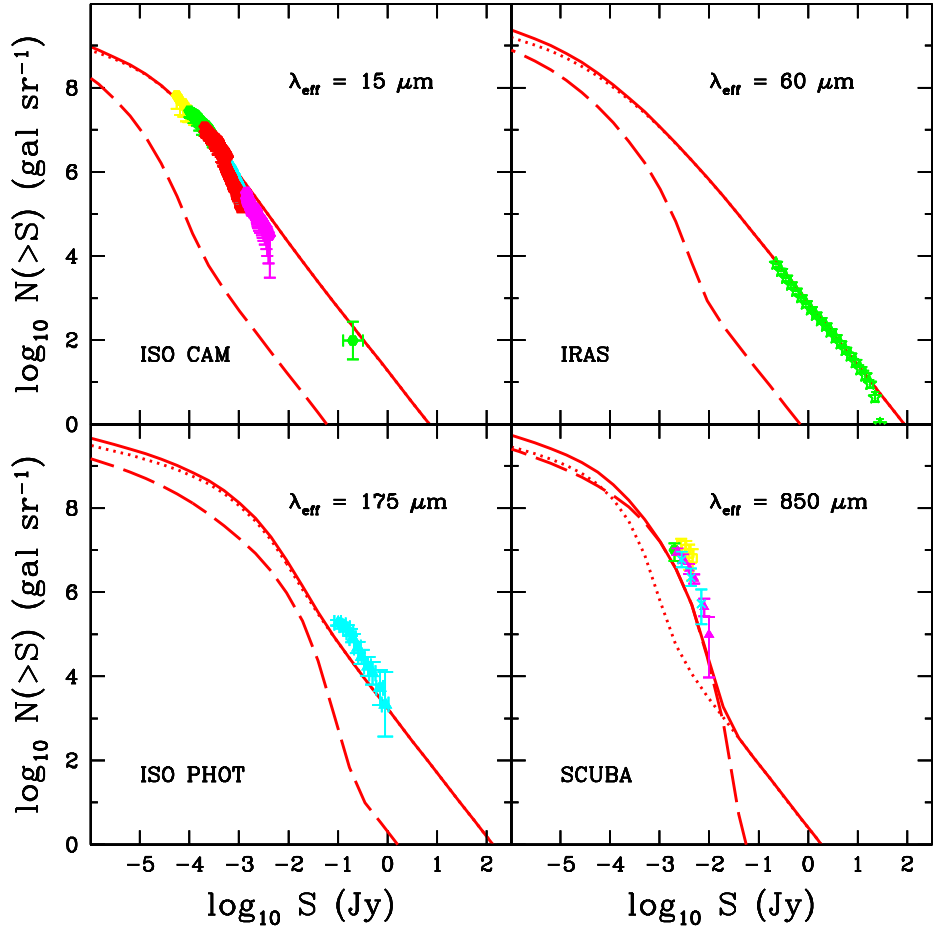}
\caption{Infrared/submm counts from the analytic model described in the text.
Coding for the lines is dotted: late type galaxies (spirals, irregulars); dashed: 
early type (ellipticals, S0s); and solid: sum of both contributions. 
Data are from Elbaz et al. (1999) (15 $\mu$m),
Kawara et al. (1998) and Puget et al. (1999) (175 $\mu$m), Smail et al. (1997), 
Eales et al. (1999) and Barger et al. (1999a) (850$\mu$m).}
\end{figure}

Finally, we point out that the major improvement with respect to 
Devriendt \& Guiderdoni (2000) is the way we model starbursts, by using a physically motivated  
merger model which allows us to differentiate between spirals and ellipticals.
This difference was established in an {\em ad-hoc} way in the model proposed by these authors,
where they were simply turning a (redshift-dependent) fraction of the spiral population into 
ellipticals. Therefore we focus on this issue in the next section.

\section{Spheroids Versus Disks}

A central issue in galaxy formation is to account for the Hubble sequence of galaxies.
The consensus view is that ellipticals form by rare major mergers and that disks form by
prolonged infall, which is equivalent to a sequence of minor mergers.  In the analytical
approach used here, we will use an approximate fit to the collision cross-section to estimate
energy exchange $\Delta$E incurred in a tidal interactions and mergers. We refer the 
interested reader to Balland et al. (1998) for a description of how to implement such 
a calculation within the Press-Schechter formalism.  

We then proceed to identify late type galaxies (spirals/irregulars) as undergoing low $\Delta$E, 
and take early types (ellipticals/S0s) to be defined by high/inter-mediate $\Delta$E. 
We define $\Delta$E by integration over galaxy masses 
and peculiar velocities with number density given by the Press-Schechter formulation.  
Normalisation of the threshold values of $\Delta$E is achieved by fitting the 
morphology-density (and also the morphology-cluster radius) relation for rich 
cluster galaxies. We adopt in the following $\Delta$E = 0.003 as the threshold between
late types and early types, which is the threshold used by Balland et al. (1998) for 
distinguishing between spirals and S0s.
The fraction of elliptical, S0 and spiral galaxies can
now be predicted as a function of redshift and of mean local density.  Field spirals 
typically form at $z = 1 - 2$, SOs at $z = 2 - 3$, and ellipticals at $z = 3 - 4$. 

\begin{figure}
\plotone{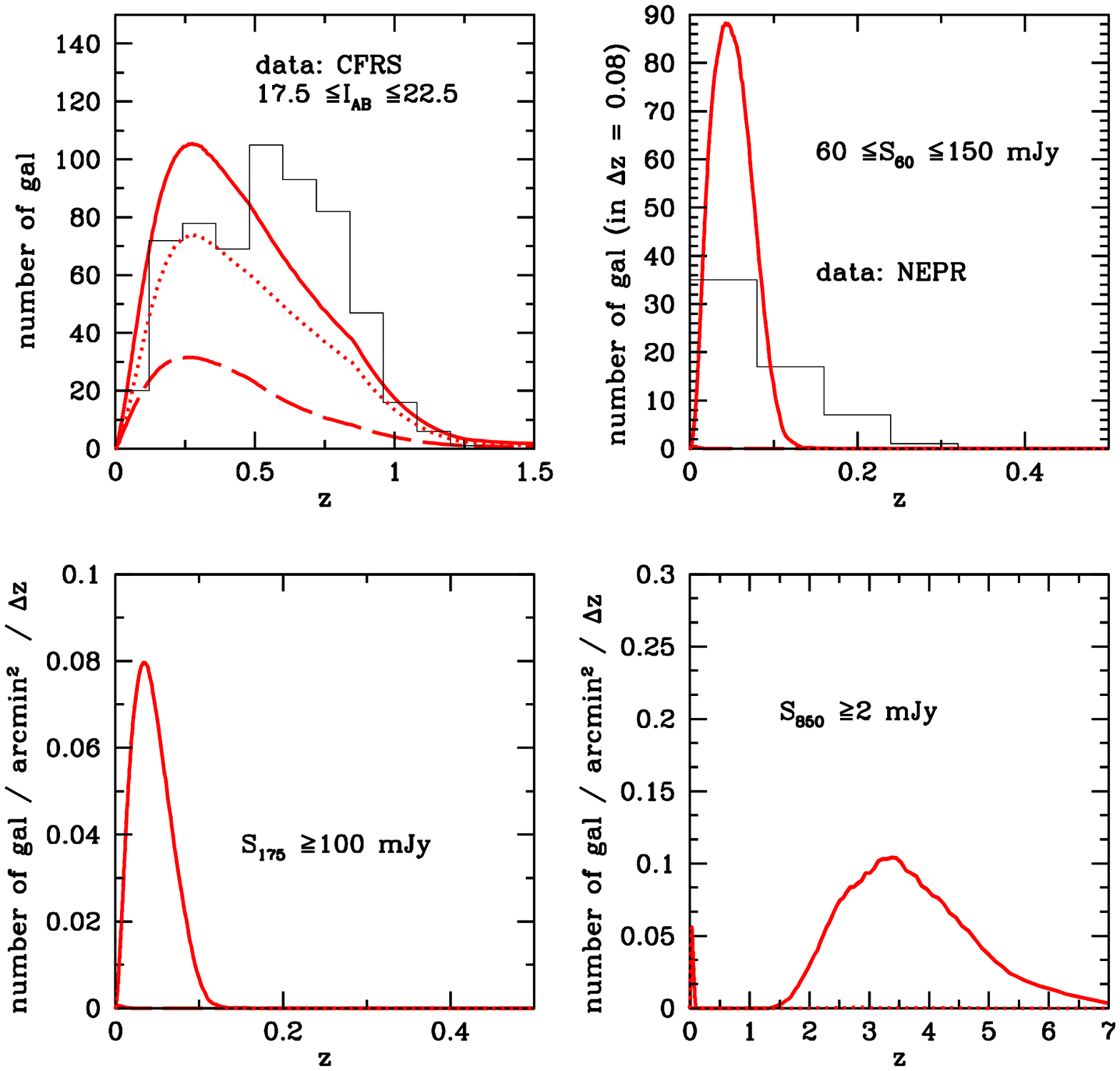}
\caption{Redshift distributions of the sources in our analytic model.
Coding for the lines is dotted: late type galaxies (spirals, irregulars); dashed: 
early type (ellipticals, S0s); and solid: sum of both contributions. 
Data are from Crampton et al. (1995) 
(Canada--France Redshift Survey), and Ashby et al. (1996) 
(North Ecliptic Pole region). Curves have been renormalised wherever data are shown.}
\end{figure}

Depending on the previously identified morphological type, we then decide which galaxies will 
undergo an ``obscure starburst'' phase. More specifically, every galaxy identified as an early 
type galaxy is supposed to go through a LIRG/ULIRG phase, whose intensity and duration are 
essentially controlled by the amount of gas available for star formation. 
In practice, this dark phase is modelled by setting our three key parameters to 
$\beta = 1$, $f_c = 1$ and $\epsilon_{SN} = 0.5$, corresponding to high star formation efficiency,
high dust opacity, and high feedback efficiency respectively.

\section{results}

The first impression is one of a fair overall agreement between model and data, as can be seen 
from figures 1 and 2.  
Looking in  more detail, one realises that (except for the faint counts in the near IR bands), 
late type galaxies dominate over early types (compare the dotted curves with the dashed
ones in each panel). This domination extends down to the far-IR, with the 
late-type galaxies still dominating the 175 microns ISOPHOT counts. 
At longer wavelengths, however, there is a dramatic change: the early-type galaxies completely 
swamp the contribution from late types. Indeed, one can see on the bottom right 
panel of figure 2 that the vast majority ($\approx$ 90 \%) of the SCUBA sources are 
classified as early types in our model.
The reason for such a change of behaviour lies in the well known negative 
k-correction, which makes galaxies of the same bolometric luminosity be as bright for the observer
at redshift 5 as at redshift 0.5. 
This is only important in the submm (here for SCUBA at 850 microns), because the peak 
emissivity of 
dust in these sources' own reference frame is between 60 and 100 microns. Therefore,
as our S0s/ellipticals form at $z > 2$, these maxima of emission must be redshifted at wavelengths 
greater than 180 and 300 microns respectively.     

Although this general agreement of our predicted counts with the multi-wavelength data 
seems quite impressive, there are several caveats.
At 15 microns, for example, one would say that we match the integral counts fairly well 
(upper left panel of figure 2). But looking more closely, we
cannot reproduce the change of slope seen in the ISOCAM differential counts. 
There are at least a couple of reasons why this could happen. 
First, the SEDs of the ISOCAM galaxies might be different in the mid-IR from the ones used 
here as a template, which are based on IRAS observations of the local universe. 
This could arise, for example, if the  grain size distribution peaks towards smaller grains.
 Second, because of 
the way interactions are modelled, each early type galaxy undergoes a starburst
after its host halo has just collapsed, and this obviously does not properly 
describe these sources 
(see figure 2), for the vast majority of them are LIRGs (not ULIRGs). 
Therefore, dynamical interactions (which are not  modeled here), triggering 
multiple milder starbursts, with time delays between them,  might provide a more realistic
 description
of these sources. 
Also, our redshift distributions seem to peak at too low redshifts especially in the I band 
where the mean redshift of the distribution is $\approx$ 0.4,
 whereas the observed peak  is more 
like 0.6 (see top left panel of figure 3). Moreover, a high redshift tail seems to be 
missing from the model (top right panel of figure 3).
Both of these difficulties (differential counts at 15 $\mu$m and redshift distributions) 
would be eased if a new component of LIRGs is included with a grain distribution whose rest 
frame emissivity peaks towards 20 $\mu$m. Furthermore, we point out that
the redshift distribution is sensitive to the choice of cosmological model: inclusion 
of $\Lambda$  would shift the peak towards higher redshifts and also  produce a high redshift tail.
Finally, we focus the reader's attention on the dramatic change occuring in the redshift 
distribution of sources when going from ISOPHOT to SCUBA, {\em i.e.} from the far-IR, to the 
submm window (bottom panels of figure 3). 
Once again, this is due to the negative k-correction boosting observed fluxes 
in the submm wave-bands.

\begin{figure}
\plotone{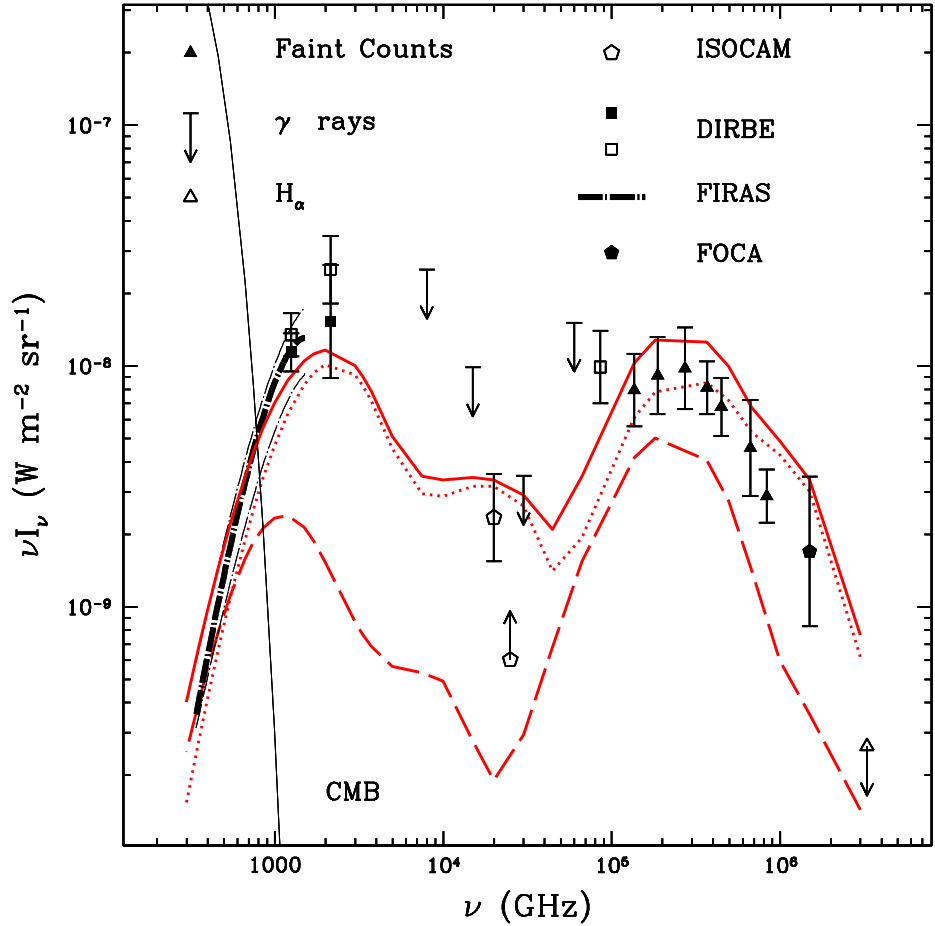}
\caption{Diffuse background from the UV to the submm as predicted from the model.
Coding for the lines is dotted: late type galaxies (spirals, irregulars); dashed: 
early type (ellipticals, S0s); and solid: sum of both contributions. The data sources 
are indicated on the figure.}
\end{figure}

\begin{figure}
\plotone{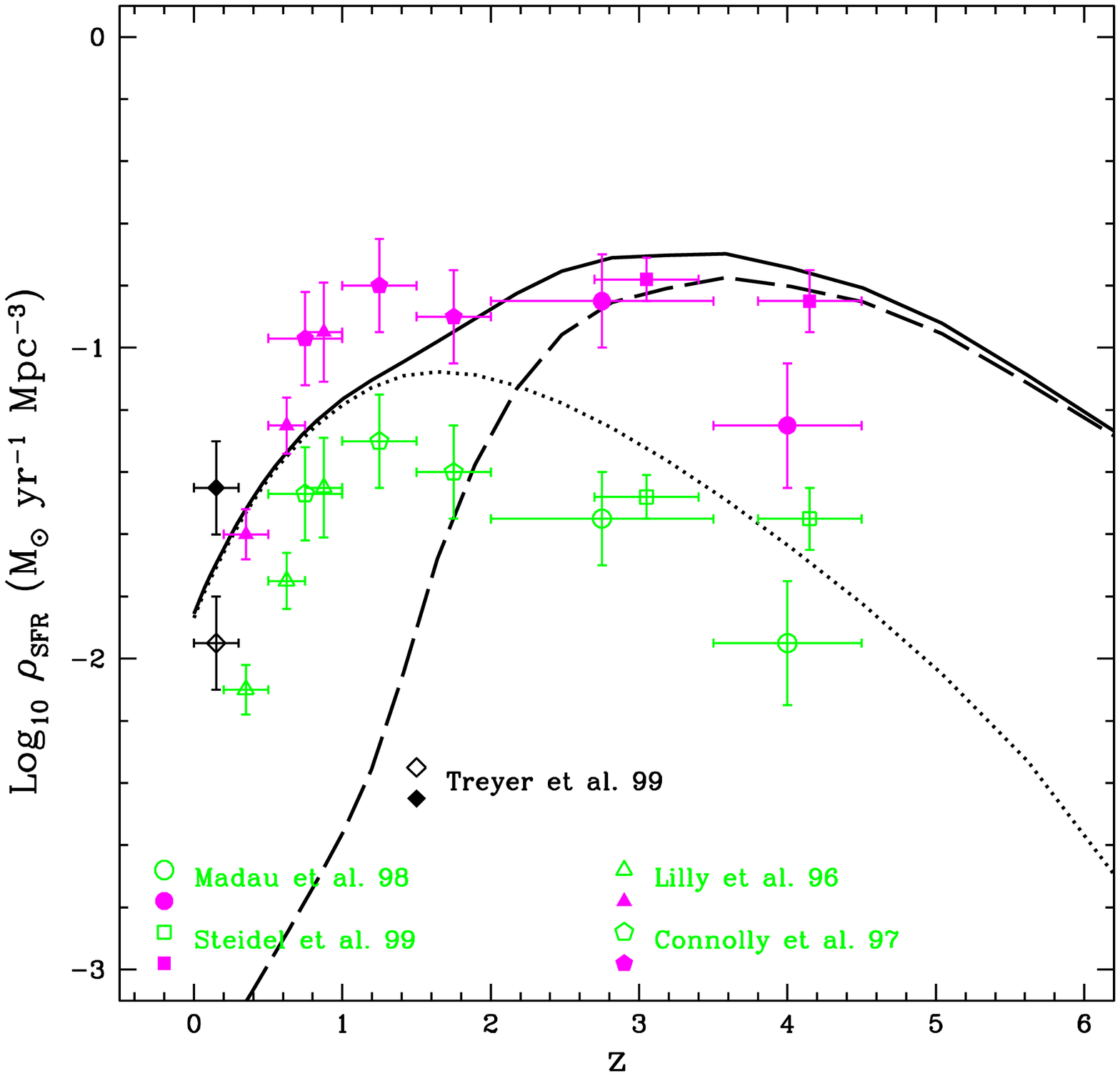}
\caption{Comoving star formation rate predicted by the model.
Coding for the lines is dotted: late type galaxies (spirals, irregulars); dashed: 
early type (ellipticals, S0s); and solid: sum of both contributions. Empty symbols give 
raw measurements and filled ones indicate tentative dust corrections (a factor of 3). 
The data sources are indicated on the figure.}
\end{figure}

The background light, as integrated from the multi-wavelength counts, seems to closely match 
the detection by Puget et al. (1996) and Hauser et al. (1998) in the far-IR/submm 
window (see figure 4). This tells us that we are  not grossly overestimating  or 
underestimating the faint counts,
and that the global galaxy luminosity budget from the  UV to the submm is 
likely to be correctly computed.
An interesting remark is that this plot clearly shows that one has to go to the 
submm, around 750 GHz (about 400 microns), to see the early-type galaxies dominate over the 
late types. At any wavelength shorter than this, quiescent galaxies, forming up to $\approx$ 
10 M$_\odot$ of stars per year dominate the total light emission (as well as the counts).

The comoving star formation rate (figure 5) is in fair agreement with the measured one,
and the reasonable amount of K-band light produced (figure 1)
makes us confident that we do not over-produce metals. Furthermore, from this perspective,
the star formation rate could be higher, provided that a fair fraction of the metals 
are ejected into the IGM. The models however seem to point to a peak in star formation around
redshift $\approx$ 3.5, sensibly higher than what is usually assumed, although current data 
are compatible with a flat star formation rate at redshifts $>$ 1. 
We argue that the amount of star formation  occurring in dust-shrouded objects cannot be much 
greater than our model predicts in order to  avoid overestimating
 the submm diffuse background. 
Therefore we emphasize  that cosmic star formation has to decrease at redshifts $\ge$ 3.5.
Note that this result is strengthened if dust-shrouded AGNs are a major contributor 
to the submm emission, even though this does not appear  to be the case 
if X-ray characteristics are a reliable AGN monitor (see Mushotzky, this volume).

\section{Conclusions}

We have been able to develop an analytical model that enables us to
separate the contributions of disk galaxies and spheroids to the
panchromatic galaxy counts, the cosmic star formation history and the
diffuse background radiation. The quiescent star formation mode
characteristic of disks self-consistently accounts for all optical,
NIR and FIR data.  Only in the submillimetre range 
 does the starburst mode, associated with spheroid formation,
dominate. It will be particularly interesting to exploit  observations near 400 $\mu$m,
where a flux-limited survey should find comparable numbers 
of forming disks and protospheroids/ellipticals.

\end{document}